\documentclass[reprint,superscriptaddress,amsmath,amssymb,aps,prl]{revtex4-2}
\usepackage{graphicx}
\usepackage{dcolumn}
\usepackage{bm}
\usepackage{subfigure}
\usepackage{hyperref}
\usepackage{color}
\usepackage{mathrsfs}
\hypersetup{
 colorlinks=true,
 linkcolor=blue,
 filecolor=magenta,
 urlcolor=cyan,
}

\begin{document}
\preprint{APS/123-QED}\preprint{APS/123-QED}

\title{ Second-order topological insulator and fragile topology in topological circuitry simulation}
\affiliation{King Abdullah University of Science and Technology (KAUST), Physical Science and Engineering Division (PSE), Thuwal 23955-6900, Saudi Arabia.}
\affiliation{King Abdullah University of Science and Technology (KAUST), Computer, Electrical and Mathematical Science and Engineering Division (CEMSE), Thuwal 23955-6900, Saudi Arabia.}
\affiliation{King Abdullah University of Science and Technology (KAUST), Physical Science and Engineering Division (PSE), Thuwal 23955-6900, Saudi Arabia.}
\affiliation{Paderborn University, Department of Physics, Warburger Str. 100, 33098
Paderborn, Germany}
\author{Ce Shang}
\email{shang.ce@kaust.edu.sa}
\affiliation{King Abdullah University of Science and Technology (KAUST), Physical Science and Engineering Division (PSE), Thuwal 23955-6900, Saudi Arabia.}
\author{Xiaoning Zang}
\affiliation{King Abdullah University of Science and Technology (KAUST), Physical Science and Engineering Division (PSE), Thuwal 23955-6900, Saudi Arabia.}
\author{Wenlong Gao}
\affiliation{Paderborn University, Department of Physics, Warburger Str. 100, 33098
Paderborn, Germany}
\author{Udo Schwingenschl\"{o}gl}
\affiliation{King Abdullah University of Science and Technology (KAUST), Physical Science and Engineering Division (PSE), Thuwal 23955-6900, Saudi Arabia.}
\author{Aur\'elien Manchon}
\email{manchon@cinam.univ-mrs.fr}
\email{aurelien.manchon@kaust.edu.sa}
\affiliation{CINaM, Aix-Marseille University, CNRS, Marseille, France.}
\affiliation{King Abdullah University of Science and Technology (KAUST), Physical Science and Engineering Division (PSE), Thuwal 23955-6900, Saudi Arabia.}
\affiliation{King Abdullah University of Science and Technology (KAUST), Computer, Electrical and Mathematical Science and Engineering Division (CEMSE), Thuwal 23955-6900, Saudi Arabia.}

\date{\today}

\begin{abstract}
Second-order topological insulators (SOTIs) are the topological phases of matter in $d$ dimensions that manifest $(d-2)$-dimensional localized modes at the intersection of the edges. We show that SOTIs can be designed via stacked Chern insulators with opposite chiralities connected by interlayer coupling. To characterize the bulk-corner correspondence, we establish a Jacobian-transformed nested Wilson loop method and an edge theory that are applicable to a wider class of higher-order topological systems. The corresponding topological invariant admits a filling anomaly of the corner modes with fractional charges. The system manifests a fragile topological phase characterized by the absence of a Wannier gap in the Wilson loop spectrum. Furthermore, we argue that the proposed approach can be  generalized to multilayers. Our work offers perspectives for exploring and understanding higher-order topological phenomena.
\end{abstract}

\maketitle
{\em Introduction.}---Identifying and classifying topological phases of matter plays an important role in modern physics. Bulk-boundary correspondence and emergent particles with quantized charge reflect the most fundamental property of a topologically protected system \cite{RevModPhys.82.3045}. A topological insulator  behaves as an insulator in its interior (bulk) in $d$ dimensions, and  $(d-1)$-dimensional metallic surface states reside in the gap of the bulk states  \cite{RevModPhys.83.1057,RevModPhys.88.021004}. Forming a novel class of topological insulators, higher-order topological insulators (HOTIs) support gapless states in a subspace of a dimension lower than $d-1$ \cite{PhysRevB.96.245115,PhysRevLett.119.246402,PhysRevLett.119.246401,Benalcazar61,Schindler2018sc,PhysRevB.97.155305,PhysRevLett.123.256402,PhysRevLett.123.216803,PhysRevLett.124.036803,PhysRevLett.124.216601,PhysRevLett.124.166804}. In two-dimensional (2D) space, a second-order topological insulator (SOTI) combines two essential ingredients: (i) fractional corner charges in the presence of certain symmetries; (ii) topologically robust zero-dimensional (0D) corner states as adiabatic deformations of the boundary with preserved symmetry \cite{PhysRevLett.123.256402,PhysRevLett.123.216803}.

While  wavefunctions and  spectra may be similar or even identical, the topological classification can be evaluated by integration of the Berry curvature over a closed manifold \cite{PhysRevLett.49.405,RevModPhys.82.1959}  or by the winding of the Wilson loop (holonomy) \cite{PhysRevB.84.075119,PhysRevLett.107.036601,PhysRevB.89.155114}.
To characterize the boundary topology, a Wannier Hamiltonian is defined as the product of ${{U}}(2N)$ Berry connection along a noncontractible Wilson loop in the Brillouin zone (BZ), and its eigenvalues are encoded with the ${{U}}(1)$ phase of the Wannier center in an effective one-dimensional (1D) form \cite{PhysRevB.84.075119}. Crucially, the Wannier centers of occupied bands refer to Wannier bands carrying a topological invariant determining the bulk-corner correspondence. These can be investigated by a nested Wilson loop formalism \cite{Benalcazar61,PhysRevLett.121.106403,PhysRevLett.123.186401,Wieder2020,Lee2020}.

HOTIs have been simulated in a plethora of platforms, including solid state \cite{Schindler2018}, photonics \cite{Mittal2019}, and acoustics systems \cite{Xue2018,PhysRevLett.122.244301,PhysRevLett.124.206601}. In the present work, we propose to employ a bilayer topological circuitry, a remarkably flexible and tunable platform \cite{Lee2018,Imhof2018}, to realize HOTIs. The layer degree of freedom splits the space of wavefunctions into a finite number of independent copies. This constitutes a direct product relationship of the spinors with the \lq layer\rq\ and \lq sublattice\rq\ parts, and the top and bottom states can be coupled by the interlayer hopping.

In this Letter, we put forward a theoretical design of SOTIs by stacking Chern insulators with opposite chiralities coupled by interlayer hopping. We start from a bilayer Haldane model \cite{PhysRevLett.61.2015}  with the time-reversal symmetry $\mathcal{T}$ and opposite Chern numbers \cite{PhysRevLett.95.226801,PhysRevLett.95.146802} (1D Dirac-type gapless helical edge states). By introducing a $\mathcal{T}$-breaking interlayer hopping, the edge modes become gapped and 0D corner modes are obtained. However, the corner modes are protected by the product of $\mathcal{T}$ and the $C_2$ rotation symmetry, leading to a \textit{fragile topological phase} \cite{ PhysRevB.96.155105,Wieder2020}. To characterize the bulk-corner correspondence, we establish a Jacobian-transformed nested Wilson loop method  by performing a Jacobian transformation of the Cartesian coordinates that reparametrizes the BZ. The corresponding topological invariant admits a filling anomaly of the corner modes with fractional charges $|e|/2$  \cite{PhysRevB.99.245151}. We utilize the edge theory to explain the underlying physics in an intuitive manner. We also claim that the stacked Chern insulator approach can be  generalized to multilayers, where only an even number of layers supports corner modes. The gap of the 1D band decays exponentially as a function of the number of layers due to finite size effects \cite{PhysRevLett.101.246807}. Our work offers new perspectives for exploring and understanding higher-order topological phenomena.

\begin{figure}[t!]
\begin{minipage}{0.5\linewidth}
\subfigure{\includegraphics[width=4.3cm]{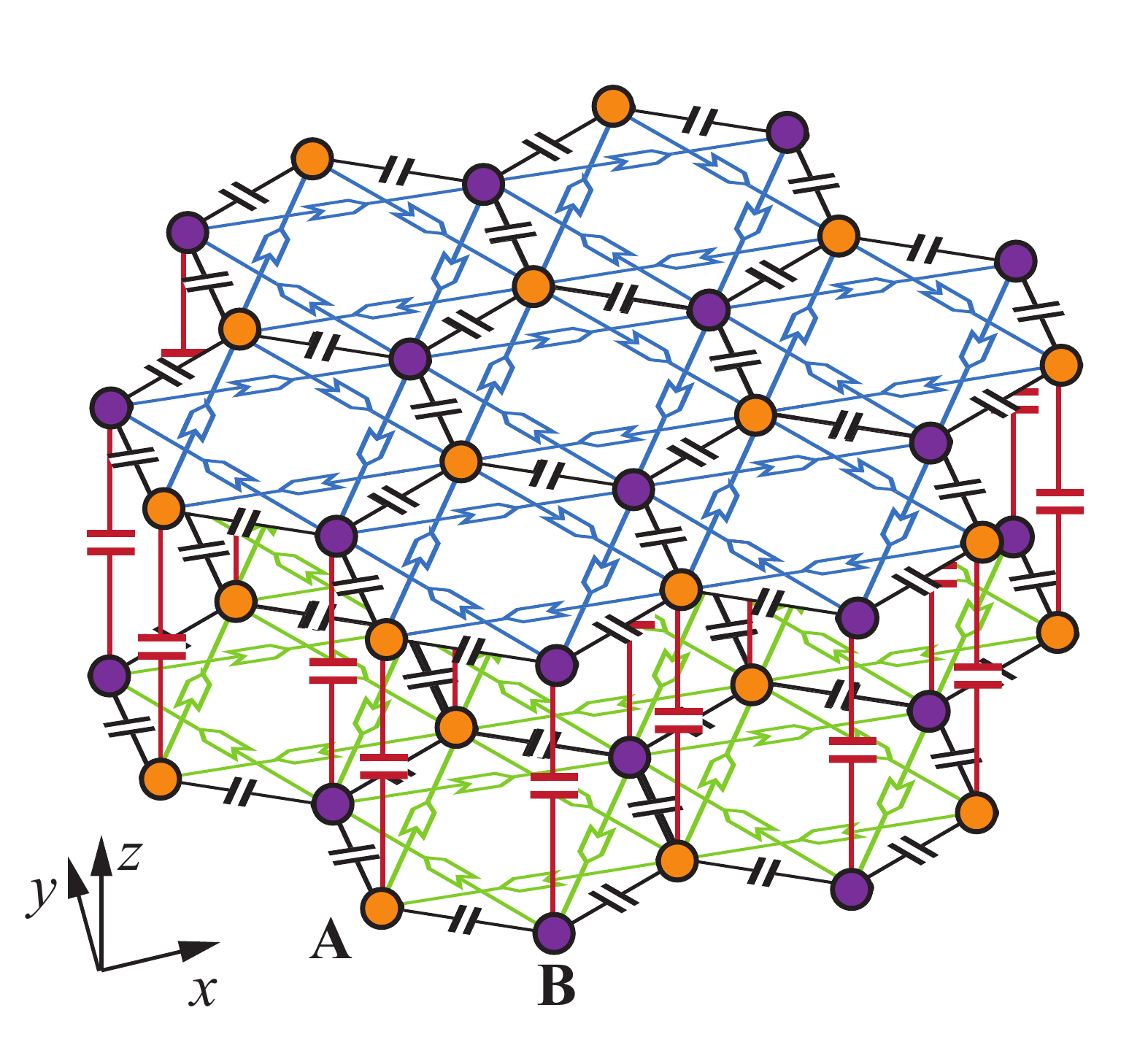}\label{fig:1a}}
\end{minipage}%
\hspace{-10mm}
\begin{minipage}{0.5\linewidth}
\subfigure{\includegraphics[width=3.7cm]{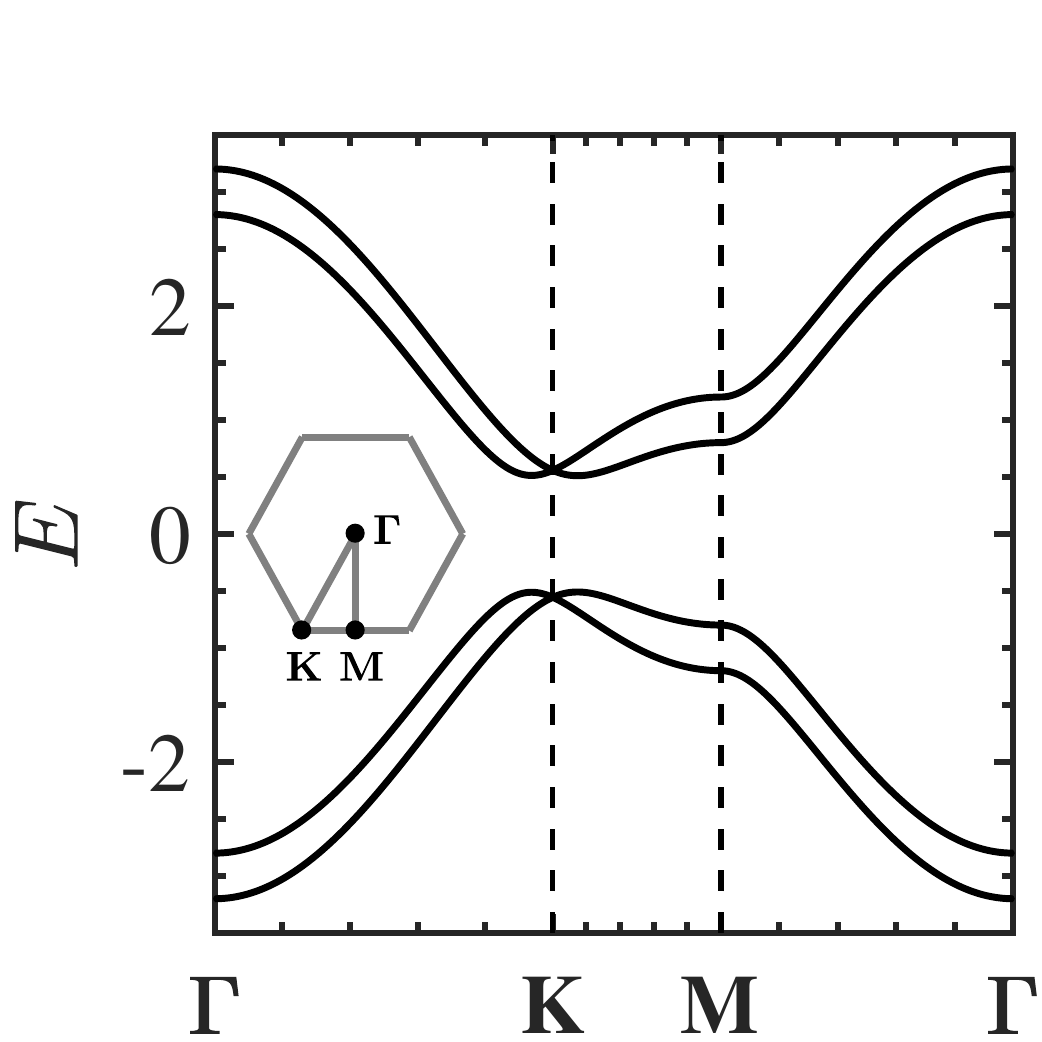}\label{fig:1b}}
\hspace{-10mm}
\end{minipage}
\begin{picture}(0,0)
\put(-230,40){(a)} \put(-105,40){(b)}
\end{picture}
\caption{(a) Schematic representation of an electrical circuit  consisting of capacitors representing the  interlayer hopping $t_1$ (black) and interlayer hopping $t_\bot$ (red), and operational amplifiers representing the spin-orbit interaction $t_2$ (blue/green) \cite{PhysRevLett.122.247702}.
(b) The energy spectrum $E$ along the high symmetry path for $t_1=1$, $t_2=0.1$ and $t_\bot=0.2$.}
\end{figure}
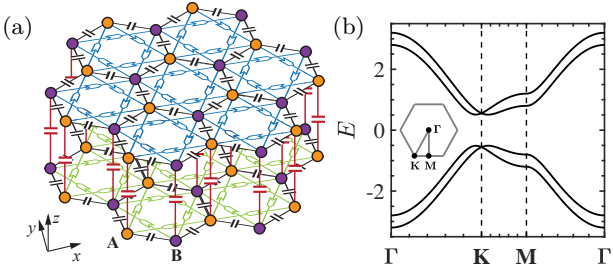

{\em Model.}--- As illustrated in Fig. \ref{fig:1a}, we investigate a honeycomb lattice layout realized by an electrical circuit \cite{PhysRevB.100.081401,PhysRevLett.122.247702}, specifically a bilayer-bipartite lattice with $A$ and $B$ sites in both of top ($\alpha$) and bottom ($\beta$) layers. The corresponding tight-binding Hamiltonian
can be written as
\begin{equation}\label{q1}
H = \sum\limits_{\left\langle {ij} \right\rangle } {{t_1}c_i^\dag {\tau _0}{c_j}}  + \sum\limits_{\left\langle {\left\langle {ij} \right\rangle } \right\rangle } {i{t_2}{v_{ij}}c_i^\dag {\tau _z}} {c_j} + \sum\limits_i {{t_ \bot }c_i^\dag {\sigma _x}} {c_i},
\end{equation}
where $c_i^\dag  = {( {c_{i,\alpha }^\dag ,c_{i,\beta }^\dag } )^T}$ is the two-component creation operator acting at the site $i$ of the $\alpha$ and $\beta$ layers.  The Pauli matrices $\sigma(\tau)_x,\sigma(\tau)_y,\sigma(\tau)_z$ and the identity matrix $\sigma(\tau)_0$ refer to the layer (sublattice) degree of freedom.  $t_1$ denotes the intralayer hopping between the nearest neighbor sites. $t_2$ emulates the spin-orbit coupling between next nearest neighbor sites, where ${v_{ij}} =  {\textbf{d}_i} \times {\textbf{d}_j}/|{\textbf{d}_i} \times {\textbf{d}_j}|$ depends on the orientations of the mediating bonds, where ${\textbf{d}_i}$ and ${\textbf{d}_j}$ are unit vectors connecting site $j$
and $i$. $t_\bot$ denotes the interlayer hopping between  site $i$ in different layers.

By Fourier transformation of the Hamiltonian (\ref{q1}) into momentum space, we obtain the Bloch matrix
\begin{equation}\label{q2}
{\cal{H}}(\textbf{k}) = {d_1}(\textbf{k}){\Gamma_1} + {d_{2}}(\textbf{k}){\Gamma_{2}} + {d_{3}}(\textbf{k}){\Gamma_{3}} - {t_ \bot }{\Gamma_{34}},
\end{equation}
with
\begin{equation}\nonumber
\begin{split}
{d_1}(\textbf{k}) =& {t_1}[1 + 2\cos (3a{k_y}/2)\cos (\sqrt 3 a{k_x}/2)],\\
{d_{2}}(\textbf{k}) =& 2{t_1}\sin (3a{k_y}/2)\cos (\sqrt 3 a{k_x}/2),\\
{d_{3}}(\textbf{k}) =& {t_2}[4\cos (3a{k_y}/2)\sin (\sqrt 3 a{k_y}/2) - 2\sin (\sqrt 3 a{k_x})],
\end{split}
\end{equation}
where ${\Gamma _{(1,2,3,4,5)}} = ({\sigma _0} \otimes {\tau _x},{\sigma _0} \otimes {\tau _y},{\sigma _z} \otimes {\tau _z},{\sigma _y} \otimes {\tau _z},{\sigma _x} \otimes {\tau _z})$ and $\Gamma _{ab} = [{\Gamma _a},{\Gamma _b}]/2i$ are the Clifford-algebra matrices. Due to common pseudospin $1/2$ representation in momentum space, we can regard the BZ to be that of a 2D hexagonal lattice with spin. By diagonalizing Eq. (\ref{q2}), we obtain the band structure shown in Fig. \ref{fig:1b}. Dimensionless units are used throughout this Letter.

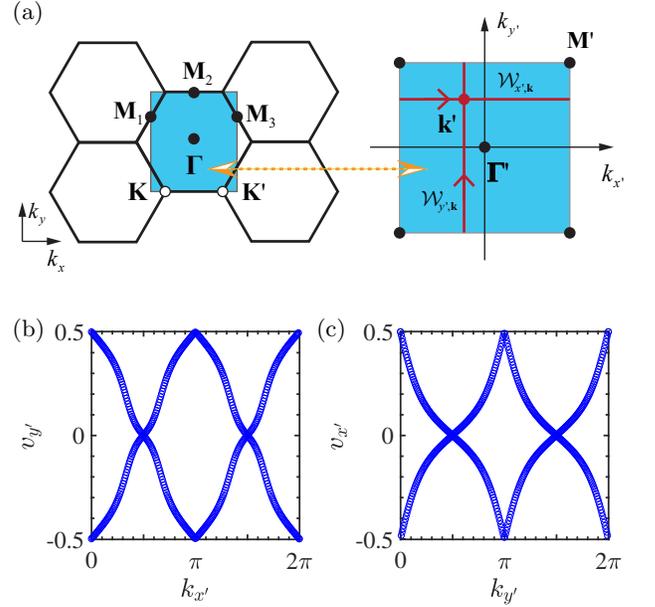
\begin{figure}[t!]
\begin{minipage}{1\linewidth}
\subfigure{\includegraphics[width=8.4cm]{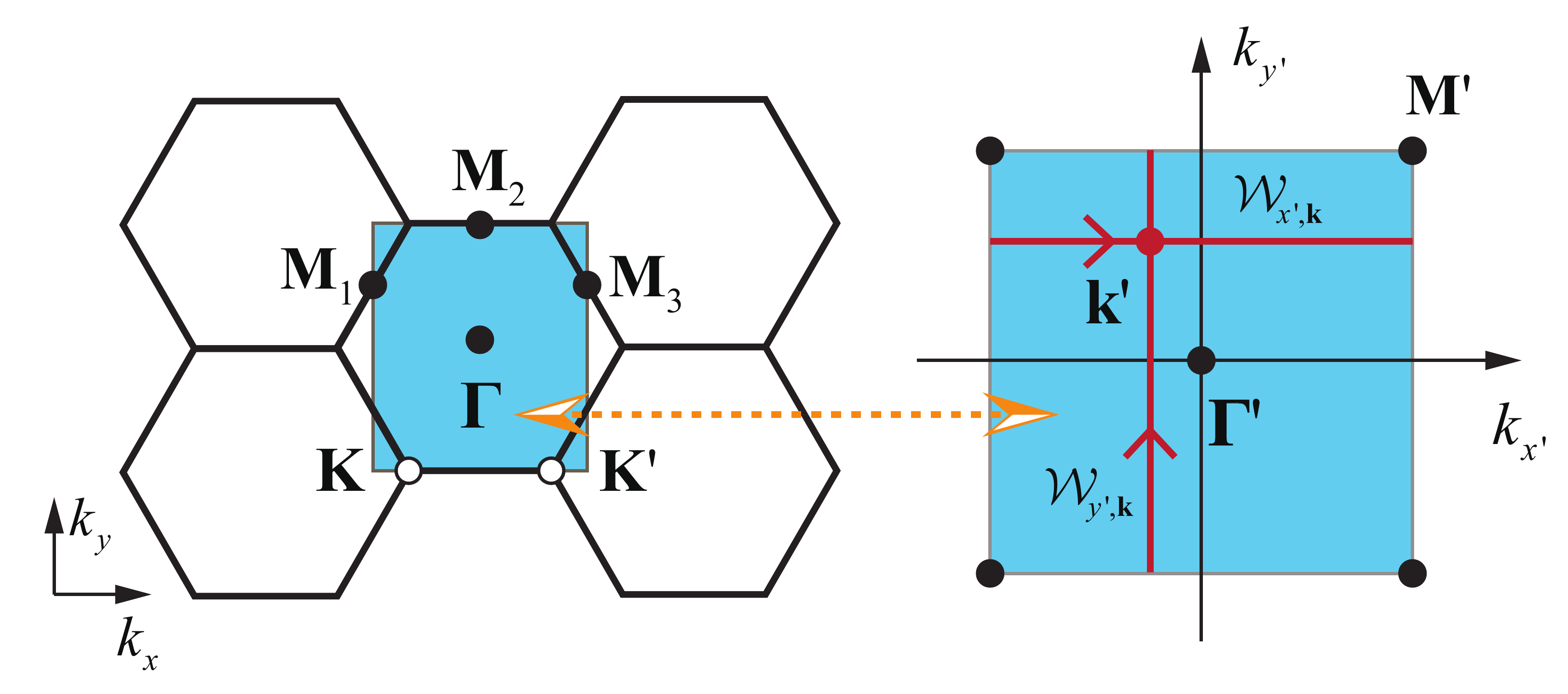}\label{fig:2a}}
\end{minipage}\\
\begin{minipage}{1\linewidth}
\subfigure{\includegraphics[width=4cm]{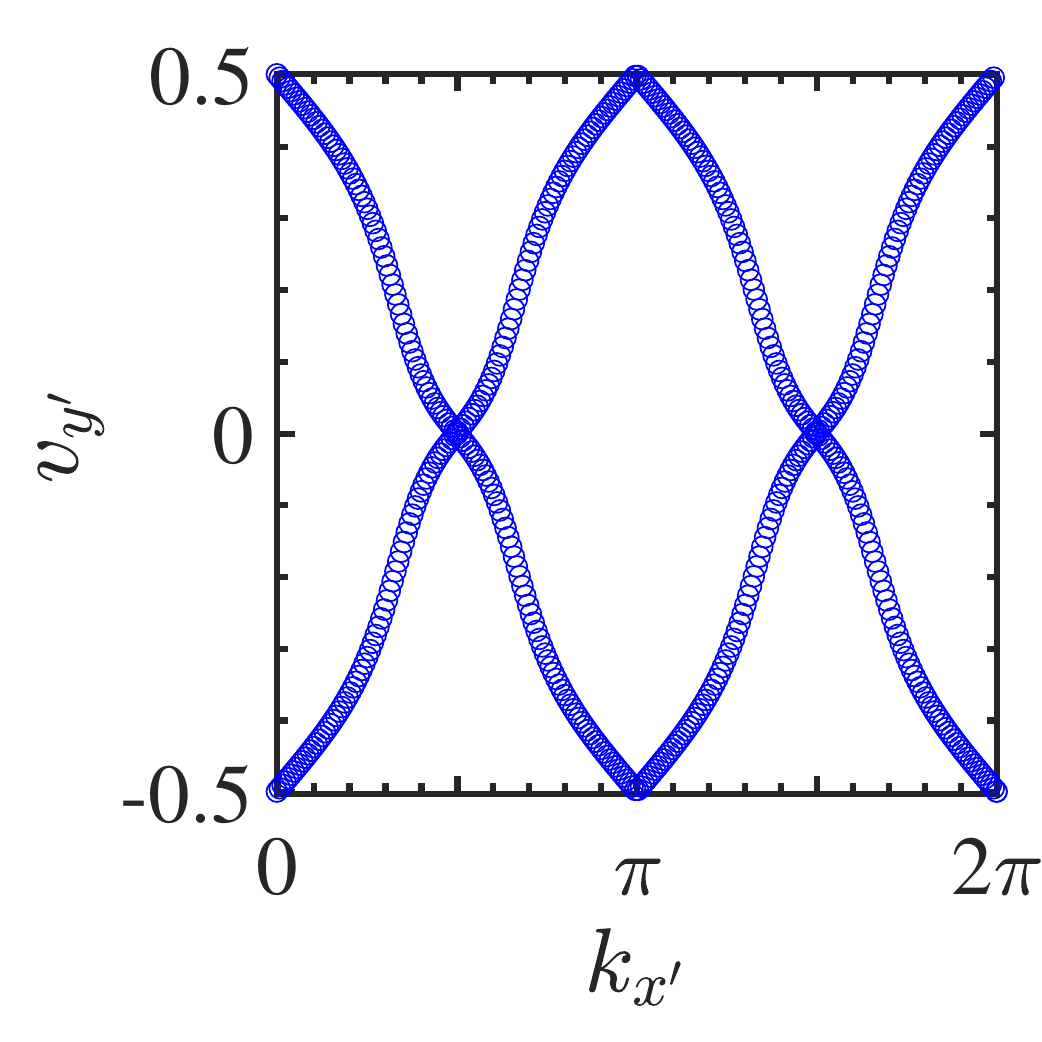}\label{fig:2b}}
\subfigure{\includegraphics[width=4cm]{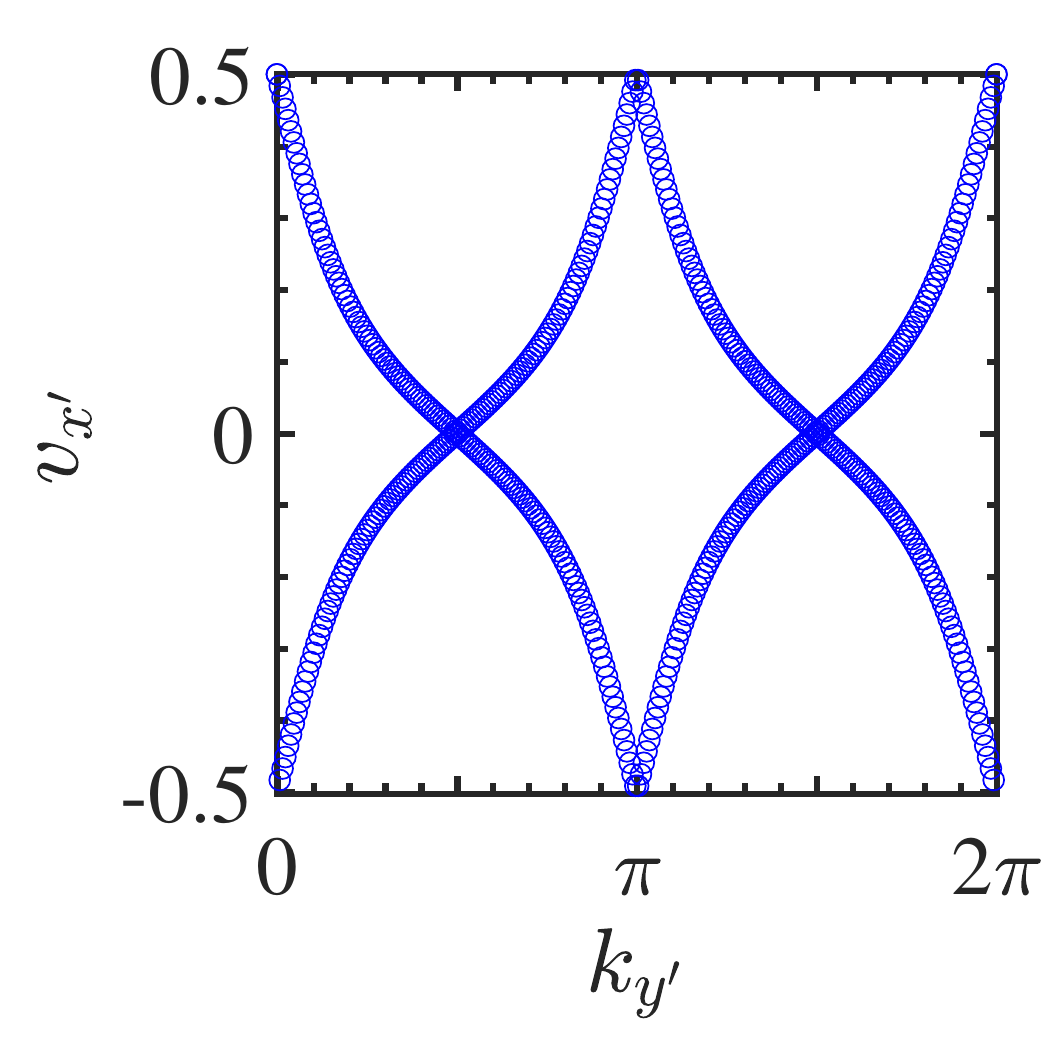}\label{fig:2c}}
\end{minipage}
\vspace{-3mm}
\begin{picture}(0,0)
\put(-115,230){(a)} \put(-115,110){(b)} \put(0,110){(c)}
\end{picture}
\caption{(a) Mapping from a honeycomb BZ to a square BZ and its corresponding with time-reversal invariant momenta, labled $\bf{\Gamma}'$ and $\bf{M}'$. ${\cal{W}}_{x'(y'),{\bf{k}}'}$ is the Wilson loop  along $k_{x'(y')}$  starting at ${\bf{k}'}=[k_{x'},k_{y'}]$. Wannier bands (b) $v_{y'}$ and (c) $v_{x'}$ constructed by Wilson loop diagonalization.}
\end{figure}

{\em Topological invariant.}---Emergence of topological corner states is the physical manifestation of a SOTI, with the topological property determined by the second Stiefel-Whitney number $w_2$ \cite{PhysRevLett.121.106403,PhysRevLett.123.186401}. $w_2$ is derived from the parity of the occupied Bloch wave functions at the time-reversal invariant momenta \cite{Lee2020} ${{{\bf{\Gamma}}} _i} \in \{ {{\bf{M}}_1},{{\bf{M}}_2},{{\bf{M}}_3},{\bf{\Gamma}}\} $ [Fig. \ref{fig:2a}] as
\begin{equation}\label{q3}
{( - 1)^{{w_2}}} = \prod\limits_{i = 1}^4 {{{( - 1)}^{[N_{\rm{occ}}^ - ({\bf{\Gamma} _i})/2]}}} ,
\end{equation}
where $[N_{\rm{occ}}^ - ({{\bf{\Gamma}} _i})/2]$ counts the number of occupied bands with odd parity.

$w_2$ also can  be determined by nested Wilson loop calculation \cite{Benalcazar61}. We introduce a Jacobian-transformed nested Wilson loop to conveniently confirm gauge-invariance. The primitive vectors of the honeycomb lattice in real and momentum space do not point along the $(x,y)$ and $(k_x ,k_y )$, respectively, as shown in Fig. \ref{fig:2a}, because the hexagonal primitive cell is topologically equivalent to a rectangular primitive cell \cite{PhysRevB.48.4442}. Hence, the transition matrix ${\mathcal{M}} =\rm{diag} [\sqrt 3 /2,3/2]$ performs the Jacobian transformation $[{k_{x'}},{k_{y'}}] = {\mathcal{M}}[{k_x},{k_y}]$, mapping the honeycomb BZ to a ${T^2} = [0,2\pi ] \times [0,2\pi ]$ torus.
The corresponding real space coordinates read $[x',y'] = {{\mathcal{M}}^{ - 1}}[x,y]$. Two manifolds with the same Euler characteristic can be continuously deformed into each other. Thus, the Wilson loop is redefined in the standardized square primitive cell as a path-ordered product of the exponential of Berry connections,
\begin{equation}\label{q4}
{{\cal{W}}_{({k_{x'}} + 2\pi ,{k_{y'}}) \leftarrow ({k_{x'}},{k_{y'}})}} = \mathop {\lim }\limits_{N \to \infty } {F_{N - 1}}{F_{N - 2}} \cdots {F_1}{F_0},
\end{equation}
where ${[{F_\ell }]_{mn}} = \langle {u_m}(\frac{{2\pi (\ell  + 1)}}{N},{k_{y'}}){\rm{|}}{u_n}(\frac{{2\pi \ell }}{N},{k_{y'}})\rangle$, $\left| u_n \right\rangle$ and $\left| u_m \right\rangle$
denote occupied Bloch functions, $N$ is the number of discrete points along the $k_{x'}$ direction, and $m, n = 1, \cdots ,N_{\rm{occ}}$. The Wilson loop eigenvalues indicates  Wannier centers $ {e^{i{v_{x'}}({k_{y'}})}}$ at given $k_{y'}$.
According to Figs. \ref{fig:2b} and \ref{fig:2c}, the Wilson loop spectra have no Wannier gap and belong to the orthogonal group $SO(2)$  \cite{PhysRevB.96.155105,PhysRevB.100.115160}, with the homotopy group ${\pi _1}[SO(2)] = \mathbb{Z}$ guaranteeing the existence
of a Stiefel-Whitney class invariant. This situation is identified as a \textit{fragile topological phase} \cite{PhysRevLett.121.126402}. The Stiefel-Whitney number can be derived from the nested Wilson loop
\begin{equation}\label{q5}
{\tilde {\cal W}_{({k_{y'}} + 2\pi ) \leftarrow ({k_{y'}})}} = \mathop {\lim }\limits_{N \to \infty } {\tilde F_{N - 1}}{\tilde F_{N - 2}} \cdots {\tilde F_1}{\tilde F_0},
\end{equation}
where ${[{{\tilde F}_\ell }]_{mn}} = \langle {v_{x',m}}(\frac{{2\pi (\ell  + 1)}}{N}){\rm{|}}{v_{x',n}}(\frac{{2\pi \ell }}{N})\rangle$. The determinant of the nested Wilson loop operator is
$\det (\tilde W)={( - 1)^{{w_2}}}$, and we can assert a nontrivial topological phase.

 {\em  Edge spectrum and topological corner states.}---
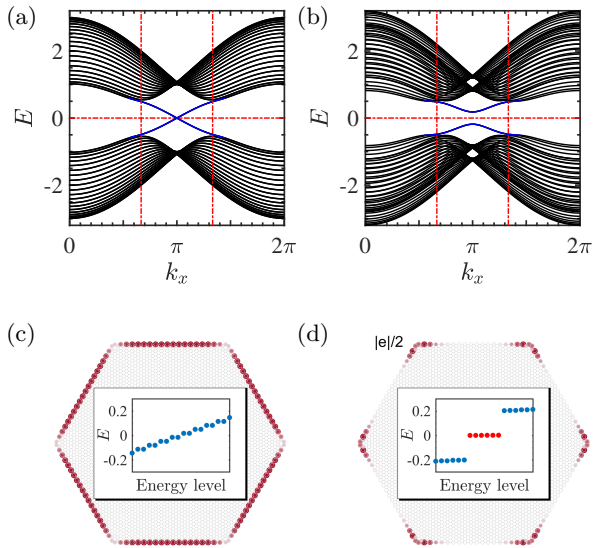
\begin{figure}[b]
\begin{minipage}{0.5\linewidth}
\subfigure{\includegraphics[width=3.8cm]{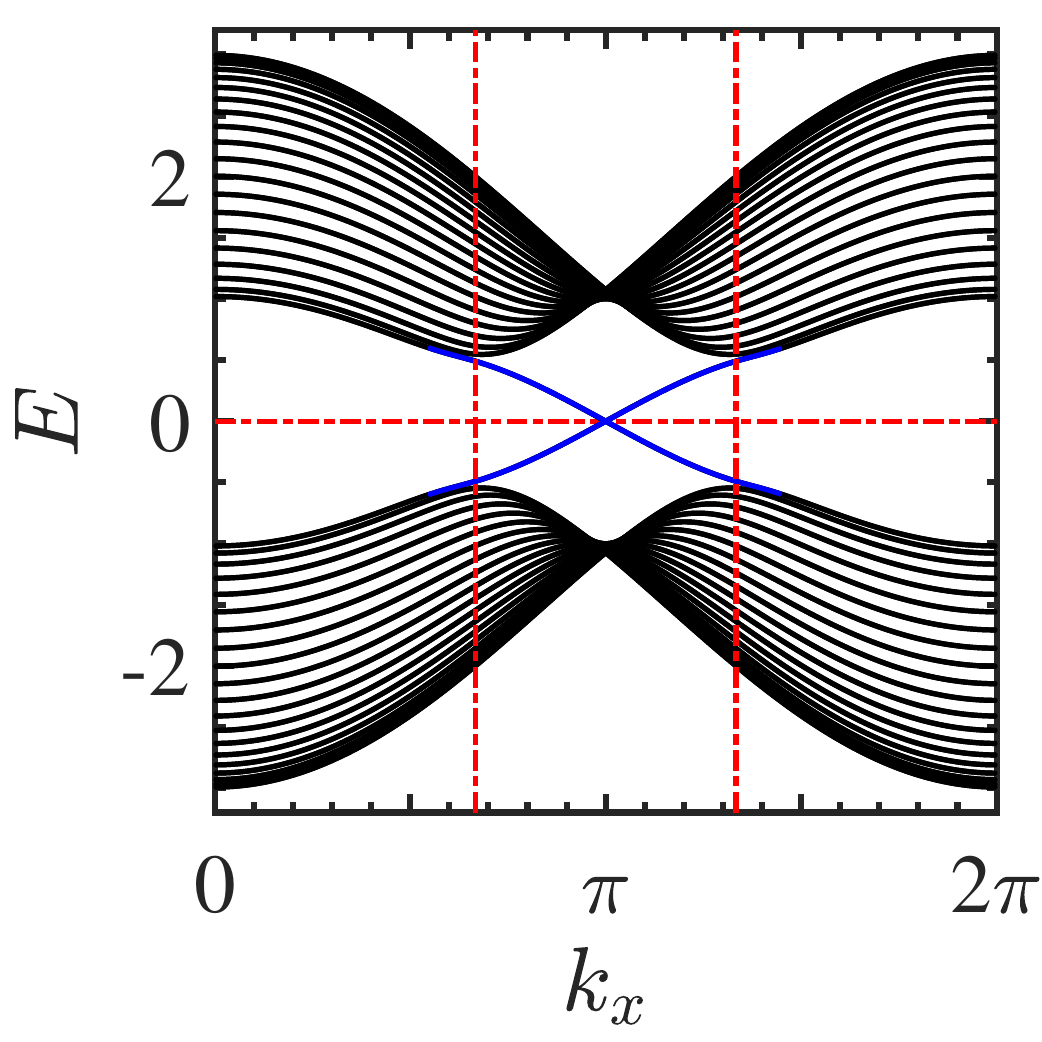}\label{fig:3a}}
\end{minipage}%
\hspace{-5mm}
\begin{minipage}{0.5\linewidth}
\subfigure{\includegraphics[width=3.8cm]{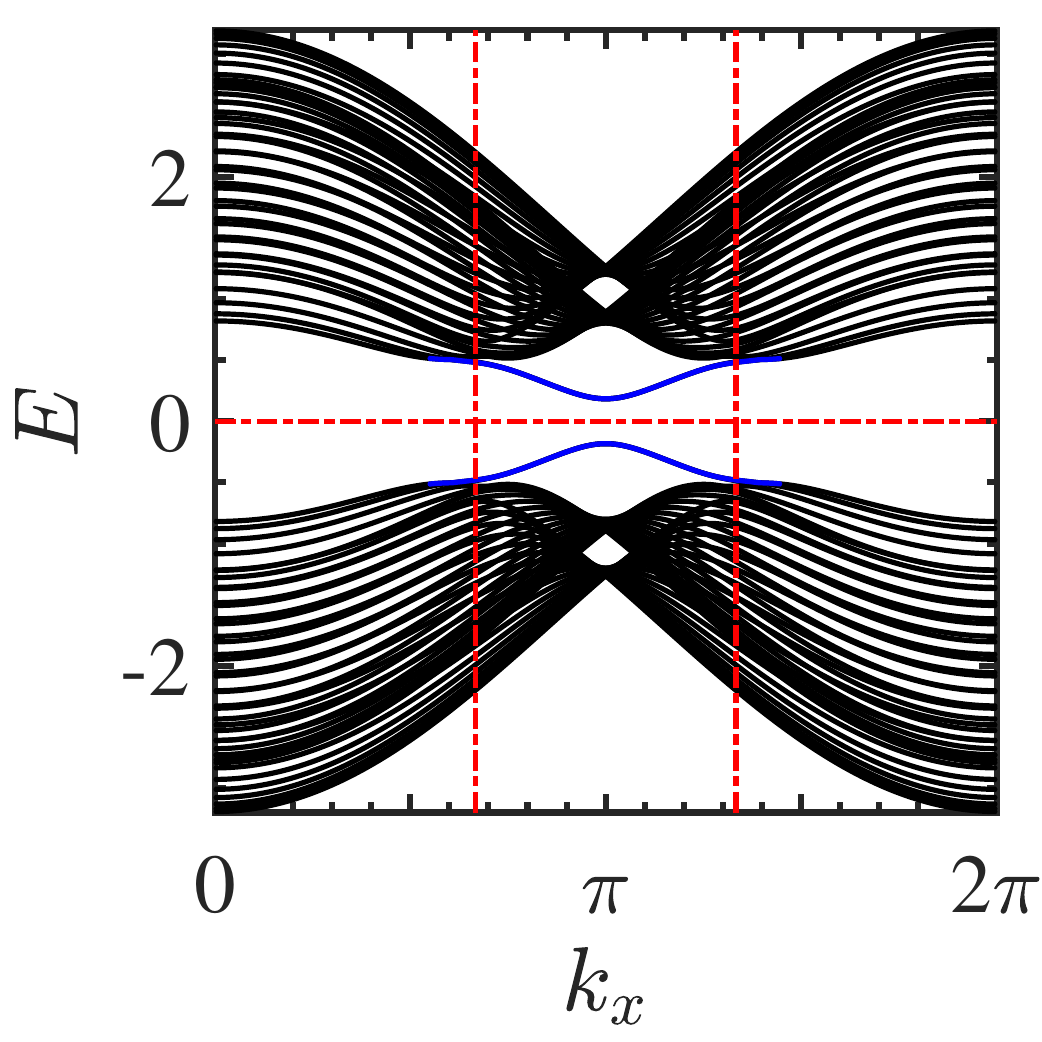}\label{fig:3b}}
\end{minipage}\\%
\vspace{3mm}
\begin{minipage}{0.5\linewidth}
\subfigure{\includegraphics[width=4.3cm]{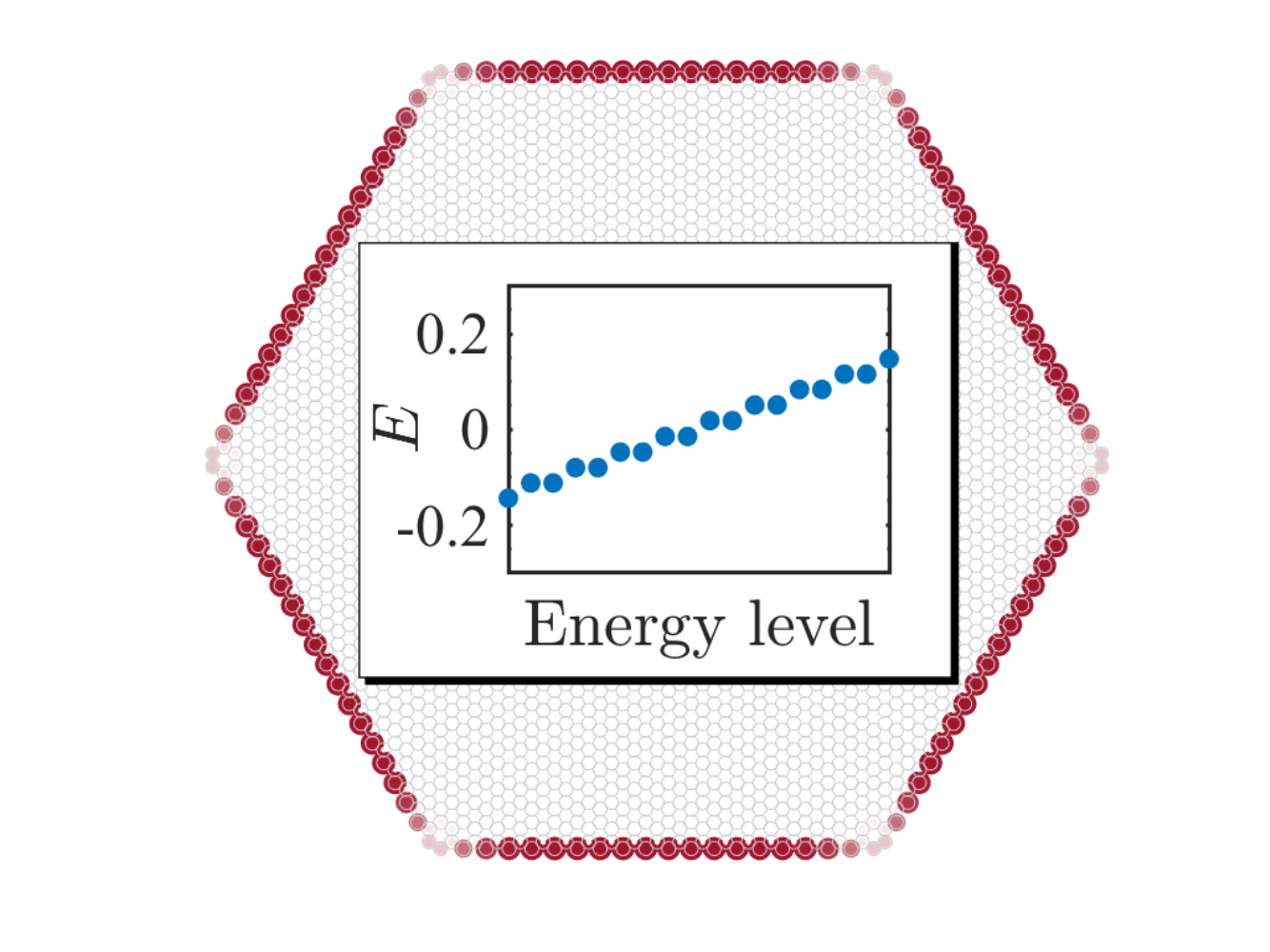}\label{fig:3c}}
\end{minipage}%
\hspace{-4mm}
\begin{minipage}{0.5\linewidth}
\subfigure{\includegraphics[width=4.3cm]{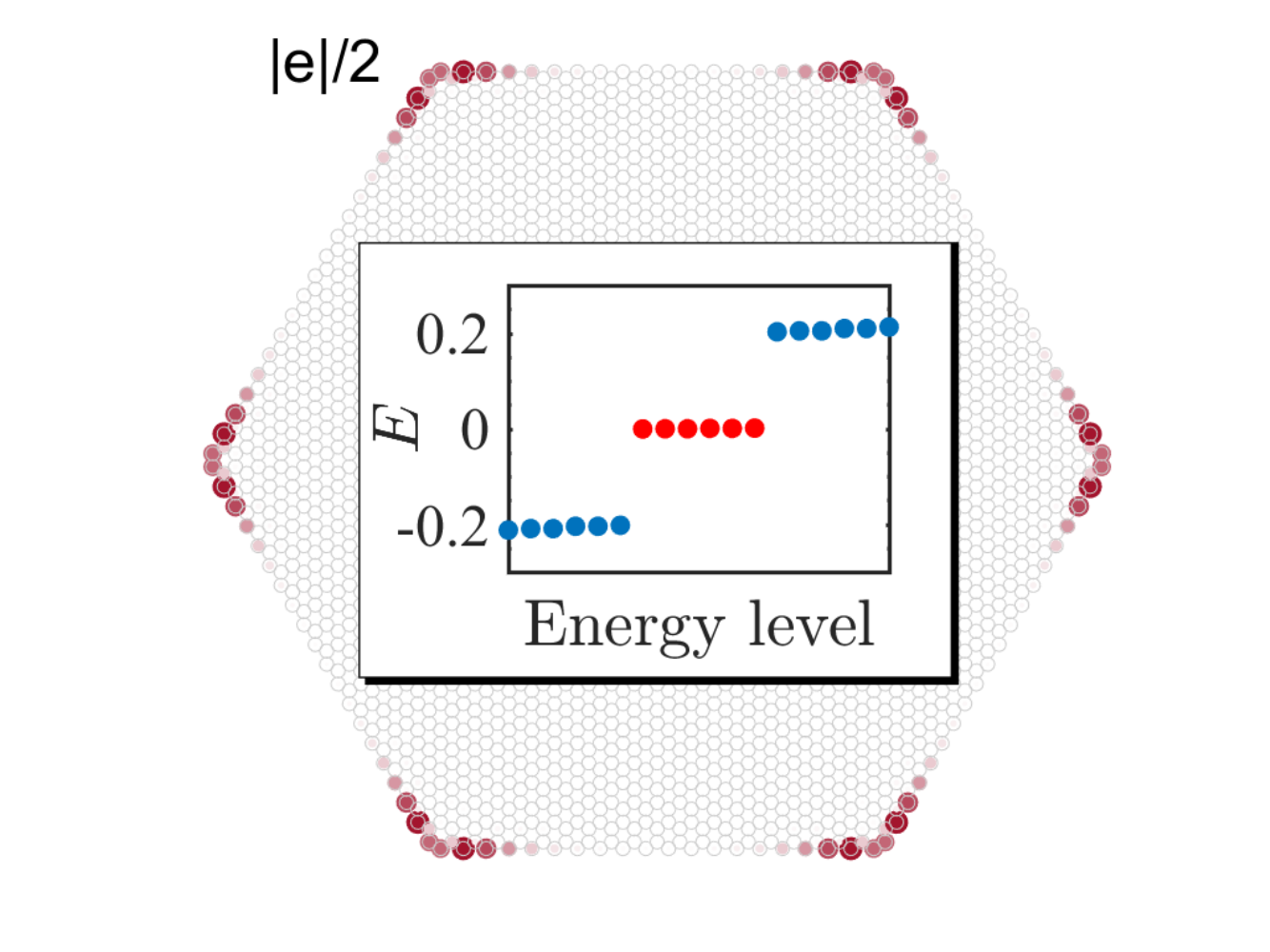}\label{fig:3d}}
\end{minipage}%
\hspace{-5mm}
\vspace{4mm}
\begin{picture}(0,0)
 \put(-225,165){(a)}\put(-115,165){(b)}
 \put(-225,45){(c)}\put(-115,45){(d)}
 \end{picture}
\caption{Band spectra  in a quasi-1D bilayer strip consisting of 10 unit cells (a) without and (b) with interlayer hopping ($t_\bot=0.2$), where, $t_1=1$ and $t_2=0.1$.  (c/d) Wavefunction for (a/b). The insets show the corresponding edge spectra.  }
\end{figure}
 We now investigate  the edge spectrum and  the physical properties of topological corner states. In a zigzag terminated ribbon with open boundary condition, without $t_{\bot}$, a pair of helical gapless edge modes forms a Dirac-like band crossing projected by $\mathcal{T}$ symmetry, see Fig. \ref{fig:3a}. Figure \ref{fig:3c} shows for a hexagonal-shaped flake with open boundary condition that the wavefuntion is localized at the edges, and the spectrum given as inset is quasicontinuous as expected for quantum spin Hall edge states. With $t_{\bot}$, the  $\mathcal{T}$ symmetry is broken and the edge modes become gapped, see Fig. \ref{fig:3b}. Due to the hexagonal geometry,
 the spectrum given as inset in Fig. \ref{fig:3d} shows
 six zero modes. The spatial distribution of the wavefunction is evaluated by the local density of states $\rho ({\bf{r}}) =  - {\pi ^{ - 1}}{{\rm{Im}}[i{0^ + }{\mathbb{I}} - H({\bf{r}})]^{ - 1}}$ at zero energy, where $0^ + $ is a small broadening. When gapped edge modes encounter one another at the corners, the $C_6$ symmetry enforces  6-fold degenerate localized corner states [Fig. \ref{fig:3d}].  The fractional charges in the six symmetric sectors are given by $Q_6=\frac{e}{2}{w_2}$ \cite{PhysRevB.99.245151}.

{\em Edge theory.}--- In order to gain intuitive understanding of SOTI phase, we construct the low-energy edge theory. The effective Hamiltonian is obtained by the lowest order expansion in the $t_\bot=0$ limit with respect to the high symmetry points $\textbf{K}$:
\begin{equation}\label{q6}
{{\cal H}_{bulk}} = \hbar {v_F}{k_x}{\Gamma _1} + \hbar {v_F}{k_y}{\Gamma _2} + m_{\rm{H}}{\Gamma _3},
\end{equation}
where $v_F=3t_1a/2$ is the Fermi velocity and $m_{\rm{H}}$ is the Haldane mass. Upon replacement ${k_{x,y}}\to-i{\partial _{x,y}}$, the real-space Dirac Hamiltonian reads
\begin{equation}\label{q7}
{ { H}_{bulk}} =  - i\hbar {v_F}{\partial _x}{\Gamma _1} - i\hbar {v_F}{\partial _y}{\Gamma _2} + m_{\rm{H}}{\Gamma _3}
\end{equation}
in the 2D semi-space $x \in ({{ - }}\infty ,\infty )$ and $y \in ({{ - }}\infty ,0]$.
Manifestly, we have $\mathcal{T} {H}_{bulk}{\mathcal{T} ^{ - 1}} = {H}_{bulk}$. Solving the Dirac equation ${H}_{bulk}\Psi ({\bf{r}})=E \Psi ({\bf{r}})$, the spinor $\Psi ({\bf{r}})$ must satisfy a boundary condition that ensures a vanishing  normal component of the particle current density,
\begin{equation}\label{q8}
{J_y}({\bf{r}}) =\langle {\Psi }({\bf{r}})|[y,{H_{bulk}}]| {\Psi ({\bf{r}})} \rangle /i\hbar  = 0.
\end{equation}
A suitable boundary condition is
\begin{equation}\label{q9}
\Psi (x,y = 0) = {\mathcal{B}} \Psi (x,y = 0),\quad {\mathcal{B}}=\Gamma_{23},
\end{equation}
where $\mathcal{T}{\mathcal{B}\mathcal{T}^{-1}}={\mathcal{B}}$ and the $\mathcal{T}$ symmetry is conserved at the edges.
Solutions with Bloch wave propagation
along the edge with  wave vector $k_x$ and decaying exponentially into the bulk with length scale $\lambda$  seeking in the form ${\Psi _{{k_x}}}({\bf{r}}) = \Phi {e^{i{k_x}x + y/\lambda }}$.
With the constraint $\Gamma_{2}\Phi =  - i\Gamma_{3}\Phi $ resulting from the boundary condition, Eq. (\ref{q9}), we have
\begin{equation}\label{q10}
\hbar {v_F}k_x{\Gamma_{1}}\Phi {{ + }}(m_{\rm{H}}- \hbar {v_F}/\lambda ){\Gamma_{3}}\Phi {{ = }}E\Phi.
\end{equation}
Setting the decay length $\lambda  = \hbar{v_F}/{m_{\rm{H}}}$, one can solve Eq. (\ref{q10}) and obtains a set of bound states $\Phi_\alpha$ \cite{PhysRevB.100.205406}. To turn on $t_{\bot}$, a mass term $m_{\rm{I}}$ is added to the Hamiltonian. Thus, in the basis $\Phi_\alpha$, we have the 1D effective Hamiltonian
\begin{equation}\label{eh}
{{\cal{H}}_{\rm{eff}}} = \hbar {v_F}{k_x}{\tau _z} + m_{\rm{I}}{\tau _x}
\end{equation}
with edge energy spectrum ${E_ \pm } = \pm \sqrt {{{\left( {\hbar {v_F}{k_x}} \right)}^2} + m_{\rm{I}}^2}$. Therefore, a 0D corner state must appear at the intersection of two edges with masses of opposite sign, corresponding to the Jackiw-Rebbi topological domain wall mode \cite{PhysRevD.13.3398}.

\begin{figure}[t]
\begin{minipage}{0.5\linewidth}
\subfigure{\includegraphics[width=3.8cm]{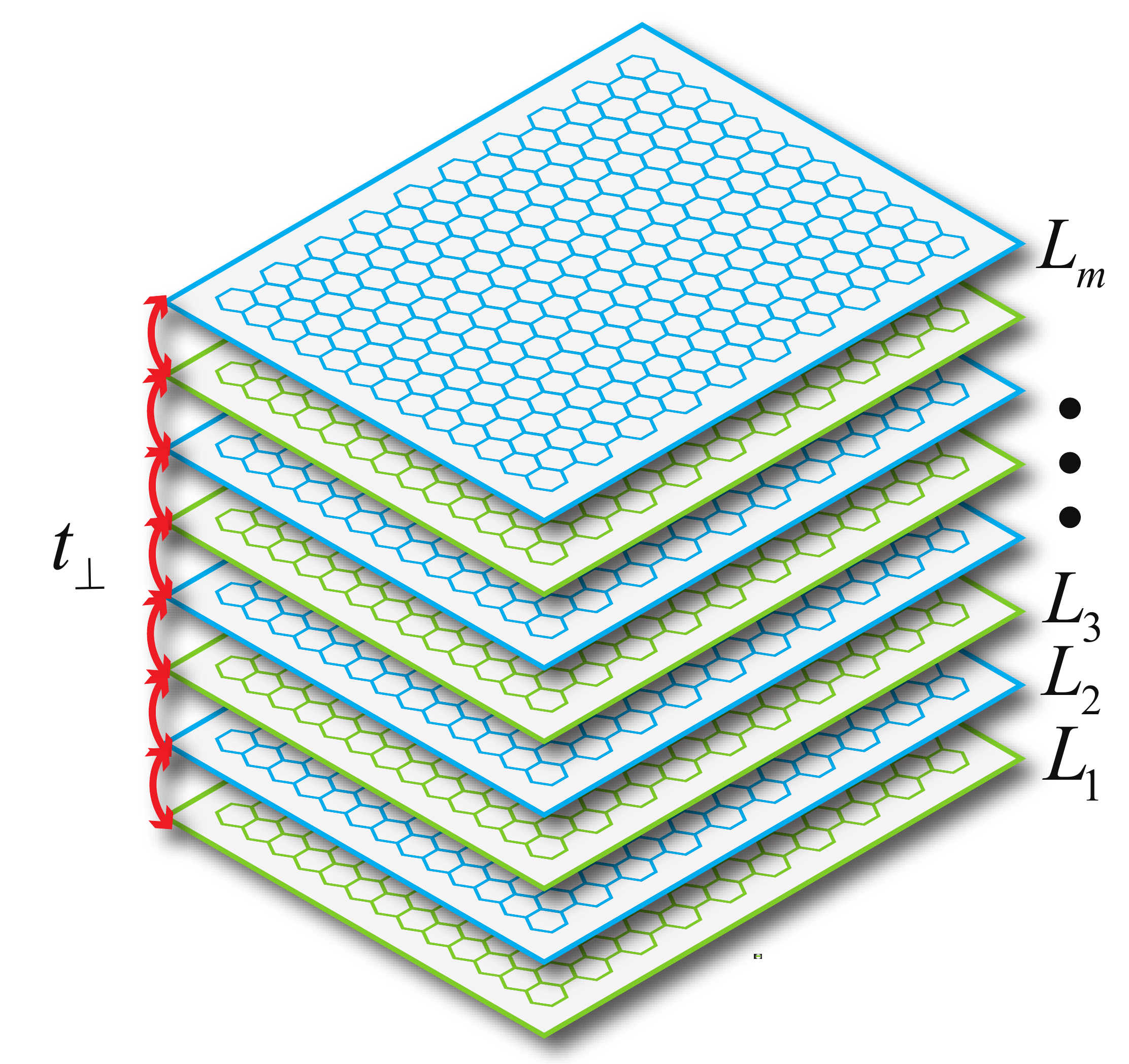}\label{fig:4a}}
\end{minipage}%
\hspace{-5mm}
\begin{minipage}{0.5\linewidth}
\subfigure{\includegraphics[width=3.8cm]{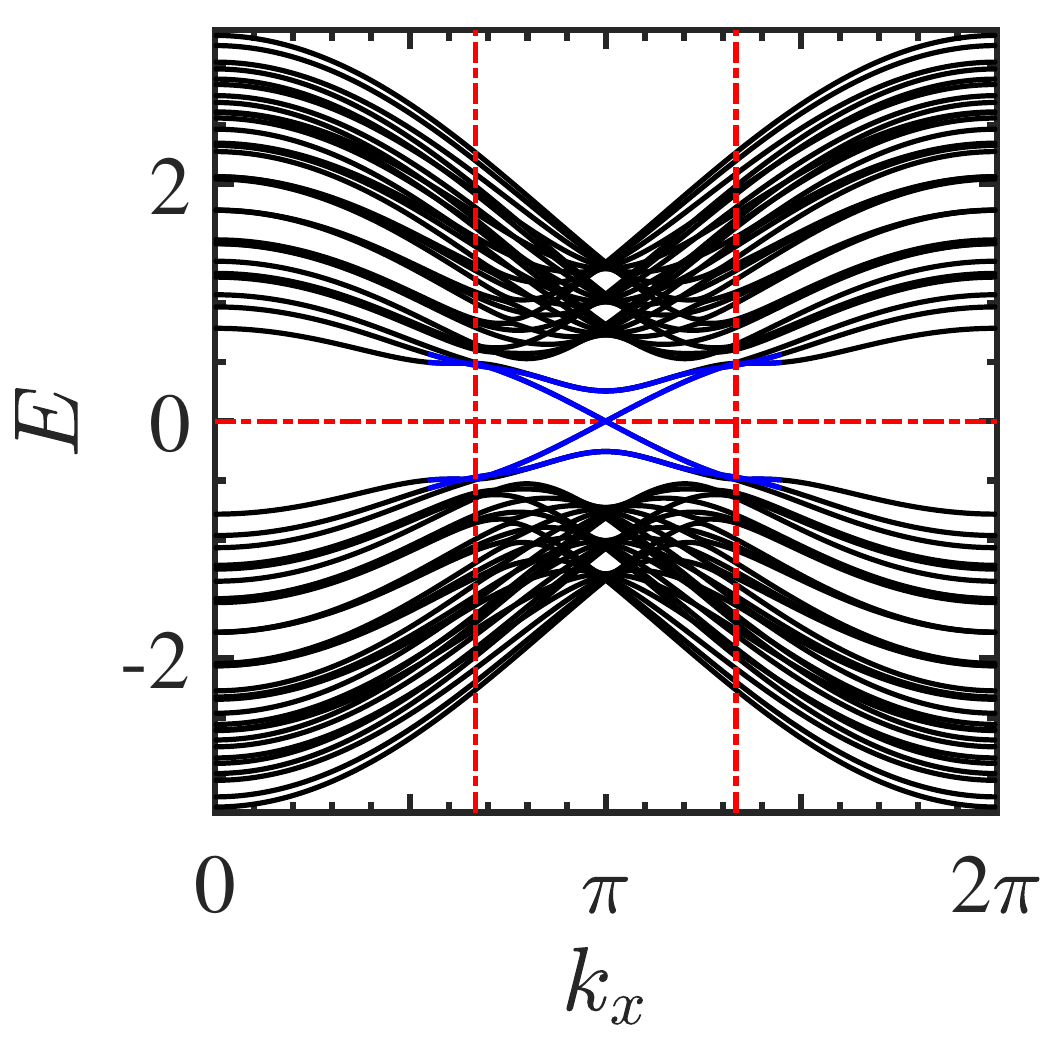}\label{fig:4b}}
\end{minipage}\\%
\vspace{-1mm}
\begin{minipage}{0.5\linewidth}
\subfigure{\includegraphics[width=3.9cm]{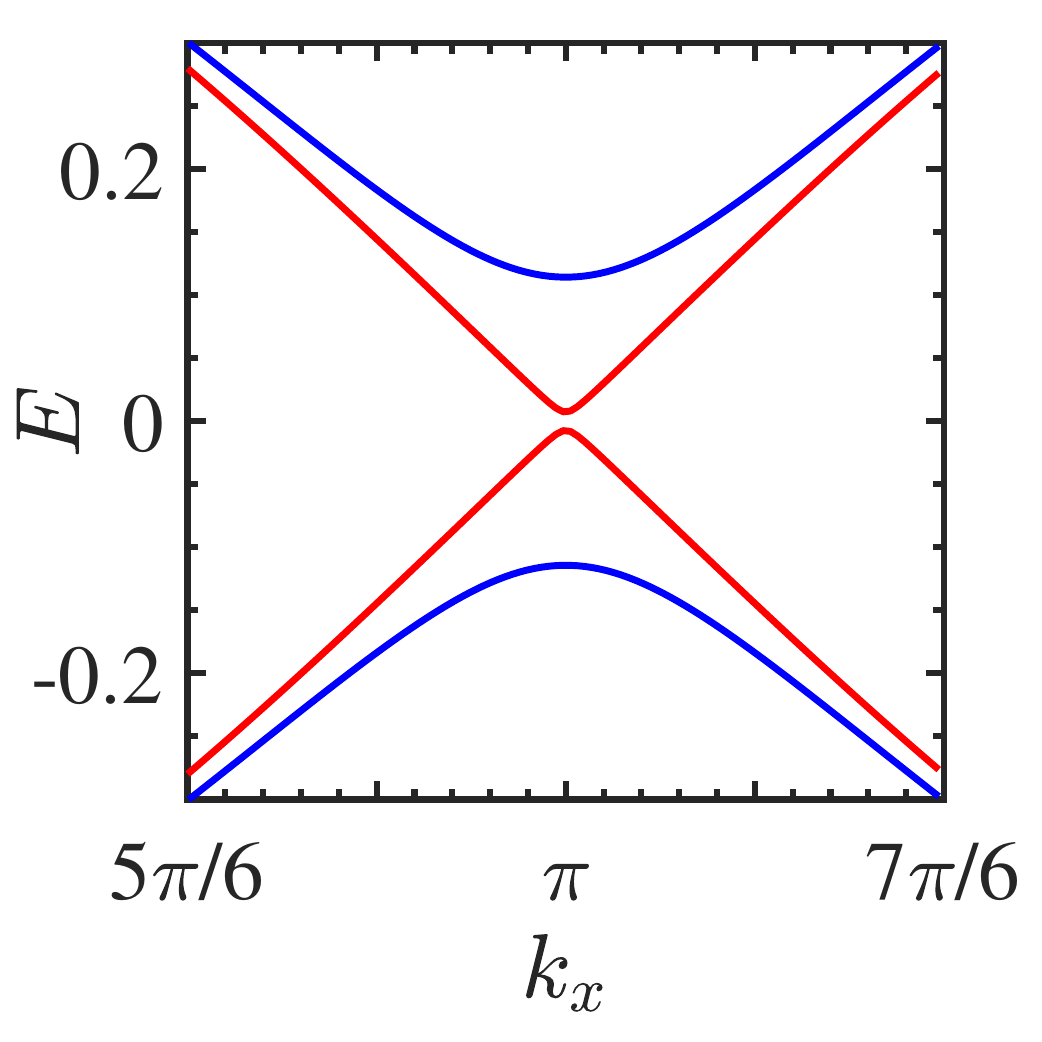}\label{fig:4c}}
\end{minipage}%
\hspace{-6mm}
\begin{minipage}{0.5\linewidth}
\subfigure{\includegraphics[width=3.9cm]{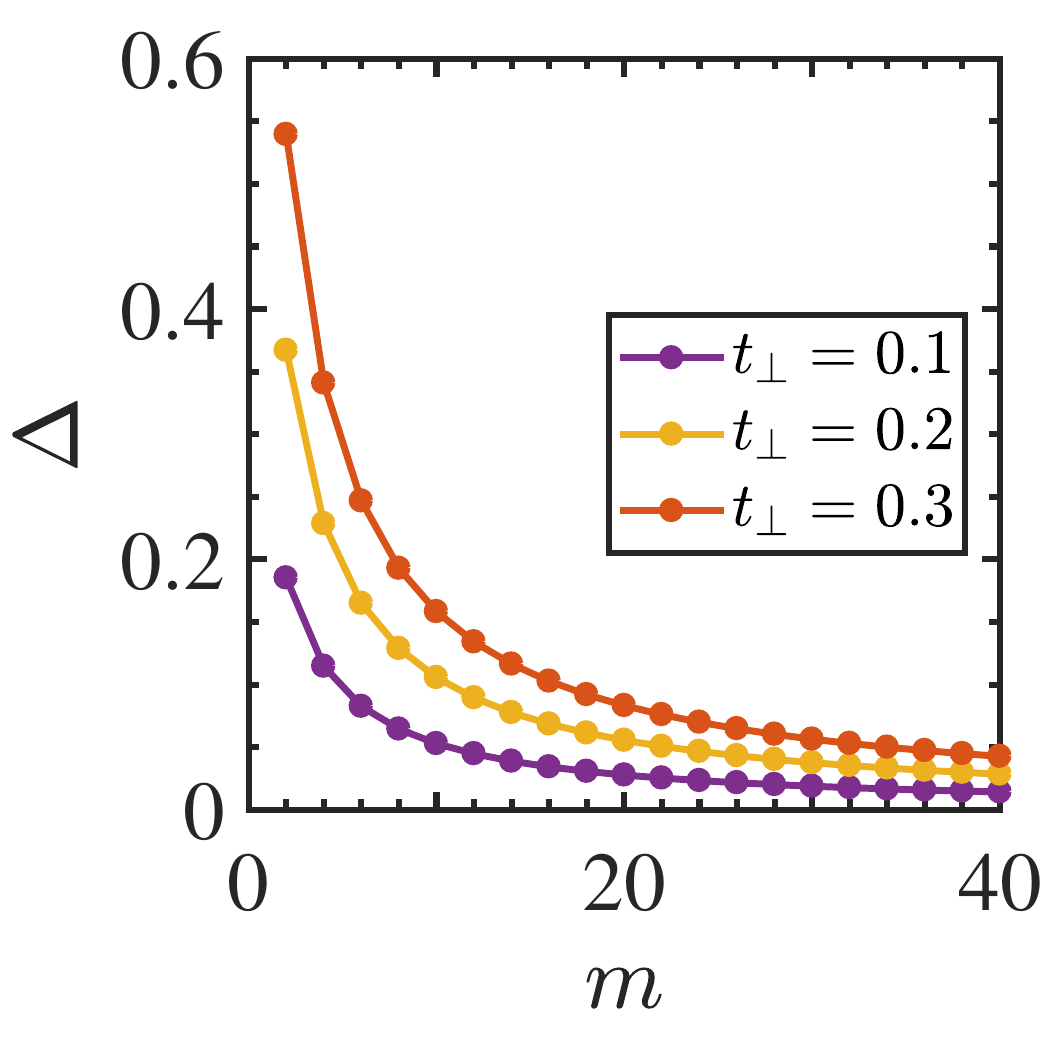}\label{fig:4d}}
\end{minipage}%
\hspace{-2mm}
\begin{picture}(0,0)
 \put(-225,165){(a)}\put(-115,165){(b)}
 \put(-225,50){(c)}\put(-115,50){(d)}
 \end{picture}
\caption{(a) Multilayer of Chern insulators with opposite chiralities connected by the interlayer hopping $t_\bot=0.2$, where $t_1=1$ and $t_2=0.1$.
(b) Band spectrum of a trilayer strip with 5 unit cells.
(c) Dispersion of edge states. (d) Energy gap  $\Delta$ as a function of the number of layers.}
\end{figure}
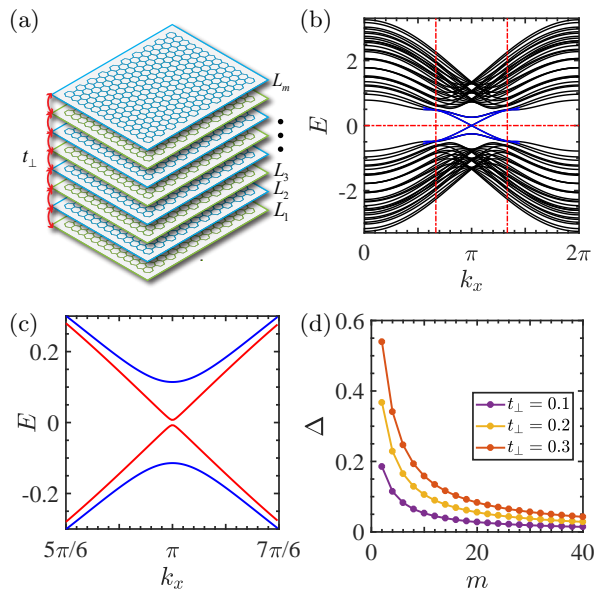

{\em Generalization to multilayers.}
The proposed approach can be generalized to  multilayers of Chern insulators, as illustrated in Fig. \ref{fig:4a}. The single-particle Hamiltonian is formed as direct sum of the Hamiltonians of the individual layers in the Hilbert space
${\mathscr{H}_m} = {\mathscr{H}_{L1}} \oplus {\mathscr{H}_{L2}} \oplus  \cdots  \oplus {\mathscr{H}_{Lm}}$.
For a state-independent interlayer coupling, we have
\begin{equation}\label{q12}
H = \sum\limits_m {C_m^\dag {C_m} \otimes {\mathscr{H}_{Lm}}}  + \sum\limits_{\left\langle {mn} \right\rangle } {C_m^\dag {C_n} \otimes {t_ \bot }\mathbb{I}} ,
\end{equation}
where $C_m^\dag$ is the creation operator acting on the whole layer $m$. We stress that only an even number of helical modes (layers) can be used for construction of a SOTI. For an odd number of layers, there will always be a single helical mode corresponding to a strong topological phase, which cannot be gapped by a local perturbation as illustrated in Fig. \ref{fig:4b}.

For an even number of layers, Fig. \ref{fig:4c} shows the lowest valence and highest conduction dispersion of edge states for $m=2$ (blue) and $m=40$ (red). The energy gap $\Delta$ between the edge states [Fig. \ref{fig:4d}] decays exponentially for increasing $m$ due to finite size effects  \cite{PhysRevLett.101.246807}. For $m\to\infty$, the gap $\Delta$ becomes tiny and the edge states are no longer well separated. This means scattering between the edge and corner states becomes possible. When the disorder or interaction scattering are introduced, the single-particle elastic scattering between the edge and corner states is no longer forbidden, implying that the corner states lose their topological protection.

{\em Summary.}---We propose and analyze a scheme to create SOTIs by stacking Chern insulators with opposite chiralities subject to interlayer coupling.
 To characterize the bulk-corner correspondence and quantify the topological invariant, we establish a Jacobian-transformed nested Wilson loop method. We demonstrate a \textit{fragile topological phase} by the absence of the Wannier gap in the Wilson loop spectrum. The topological invariant is found to admit a filling anomaly of the corner modes with fractional charges $|e|/2$. Finally, we show that the introduced approach can be  generalized to  multilayers. Our proposal can be readily realized by topological circuitry, and developing new bulk materials or van der Waals heterostructures calls for a thrilling challenge.
\bigbreak
The authors acknowledge financial support from the
King Abdullah University of Science and Technology
(KAUST).

\providecommand{\noopsort}[1]{}\providecommand{\singleletter}[1]{#1}%
%


%
%
%
%
\end{document}